\documentclass[paper=a4,abstracton,numbers=noenddot]{scrartcl}

\usepackage[english]{babel}  
\usepackage{hyperref}
\usepackage[intlimits]{amsmath}
\usepackage{units} 
\usepackage{amsmath}
\usepackage{amssymb,pifont,amsthm}
\allowdisplaybreaks[1]
\usepackage[numbers]{natbib}
\usepackage{graphicx}
\usepackage[space]{grffile} 
\usepackage{float}

\usepackage[T1]{fontenc}
\usepackage{lmodern}  		

\usepackage[scaled=.92]{helvet}
\usepackage{courier}
\usepackage[a4paper,left=2.5cm,right=2.5cm,top=3cm,bottom=2.5cm]{geometry}
\usepackage{authblk}
\usepackage{caption}

\title{Multi-level B\'{e}zier extraction of truncated hierarchical B-splines for isogeometric analysis}

\author{Andreas Grendas}	
\author{Benjamin Marussig}

\affil{Institute of Applied Mechanics, TU Graz, Technikerstraße 4, 8010 Graz, Austria}
\date{}                     %% if you don't need date to appear
\begin{document}
\maketitle

\begin{abstract}
    Multivariate B-splines and Non-uniform rational B-splines (NURBS) lack adaptivity due to their tensor product structure. Truncated hierarchical B-splines (THB-splines) provide a solution for this. THB-splines organize the parameter space into a hierarchical structure, which enables efficient approximation and representation of functions with different levels of detail. The truncation mechanism ensures the partition of unity property of B-splines and defines a more scattered set of basis functions without overlapping on the multi-level spline space. Transferring these multi-level splines into B\'{e}zier elements representation facilitates straightforward incorporation into existing finite element (FE) codes. By separating the multi-level extraction of the THB-splines from the standard B\'{e}zier extraction, a more general independent framework applicable to any sequence of nested spaces is created. The operators for the multi-level structure of THB-splines and the operators of B\'{e}zier extraction are constructed in a local approach. Adjusting the operators for the multi-level structure from an element point of view and multiplying with the B\'{e}zier extraction operators of those elements, a direct map between B\'{e}zier elements and a hierarchical structure is obtained. The presented implementation involves the use of an open-source Octave/MATLAB isogeometric analysis (IGA) code called GeoPDEs \cite{garau2018algorithms}. A basic Poisson problem is presented to investigate the performance of multi-level B\'{e}zier extraction compared to a standard THB-spline approach.
\end{abstract}

\section{Introduction}
Isogeometric analysis (IGA) \cite{hughes2005isogeometric,cottrell2009isogeometric} is a numerical method for solving partial differential equations (PDEs) by using a unified representation for both the analysis and design of structures. IGA is promising to close the gap between analysis and CAD geometries of the structures. Traditionally, engineers have used separate methods for modeling the geometry of a structure and analyzing its behavior. However, IGA breaks through this limitation by employing non-uniform rational B-splines (NURBS) functions, a standard technology employed in CAD systems. The isoparametric philosophy is invoked, and the solution space for dependent variables is represented in terms of the same functions that represent the geometry.

Refinement techniques for NURBS functions are inherently simple and do not require further communication with the CAD system. Order elevation and knot insertion are refinement techniques analogous to standard FE methods like $h$- and $p$-, and a new, higher-order methodology emerges, $k$-refinement, which combines order elevation followed by knot insertion. All subsequent meshes retain the exact geometry of the initial CAD structure. 

Due to the tensor product structure of the B-spline and NURBS functions, local refinement can not be achieved. Various approaches have been developed to overcome this limitation of tensor product structures, such as T-splines \cite{scott2011isogeometric}, LR-splines \cite{dokken2013polynomial}, and HB-splines \cite{kraft1997adaptive}. HB-splines are based on a multi-level structure with different levels of detail and appear promising to implement local refinement. Those splines can also be applied to T-splines indicating hierarchical T-splines \cite{evans2015hierarchical}. The hierarchical B-splines recently extended with the truncation mechanism, namely, THB-splines \cite{giannelli2012thb}. This work presents an adaptive isogeometric analysis framework based on THB-splines. This type of function preserves all the essential mathematical properties of standard B-splines and retains the partition of unity property in a hierarchical structure. Despite the good properties of THB-splines and their simplicity, implementing the hierarchical definition of shape functions into existing FE solvers can be rather challenging. The proposed approach for this implementation is based on a generalization of the classical B\'{e}zier extraction \cite{borden2011isogeometric}, extended into a hierarchical concept \cite{hennig2016bezier,d2018multi}. In fact, the multi-level B\'{e}zier extraction maps the multi-level hierarchical functions, and the corresponding knot spans, into a sequence of elements being equipped with a standard single-level basis.
%%%%%%%%%%%%%%%%%%%%%%%%%%%%%%
\section{Preliminaries}
In this section, some preliminary concepts need to be discussed in order to give a summarized overview of the work steps. In particular, B-splines and NURBS, the idea of B\'{e}zier extraction, and hierarchical splines, together with the truncation mechanism, are presented.
\subsection{B-splines and NURBS}
We start by introducing the concept of B-splines and NURBS functions, but further details can be found in \cite{hughes2005isogeometric}, \cite{cottrell2009isogeometric} and \cite{piegl1996nurbs}. B-splines basis functions are defined by degree $p$, and a \textbf{knot vector}:
\begin{equation}\label{one}
     \Xi = \{ \xi_1\leq...\leq \xi_{n+p+1} \},
\end{equation}
where $n$ is the number of basis functions. The {\bf Cox-de Boor} recursion formula defines the $n$ functions, by starting from degree $p=0$, up to the prescribed degree, 
\begin{equation}\label{Cox}
    N_{i,0} = \begin{cases}
                    1 ~ \text {for} ~ \xi_i ~ \leq \xi < \xi_{i+1}  \\
                    0 ~ \text {else}
                 \end{cases} 
    N_{i,p}(\xi) = {\dfrac{\xi - \xi_i}{\xi_{i+p} - \xi_i}  N_{i,p-1}(\xi)} +{\dfrac{\xi_{i+p+1} - \xi}{\xi_{i+p+1} - \xi_{i+1}}  N_{i+1,p-1}(\xi)}  . 
\end{equation}

\begin{figure}[H]
    \begin{minipage}{.49\textwidth}
    \includegraphics[scale=0.24]{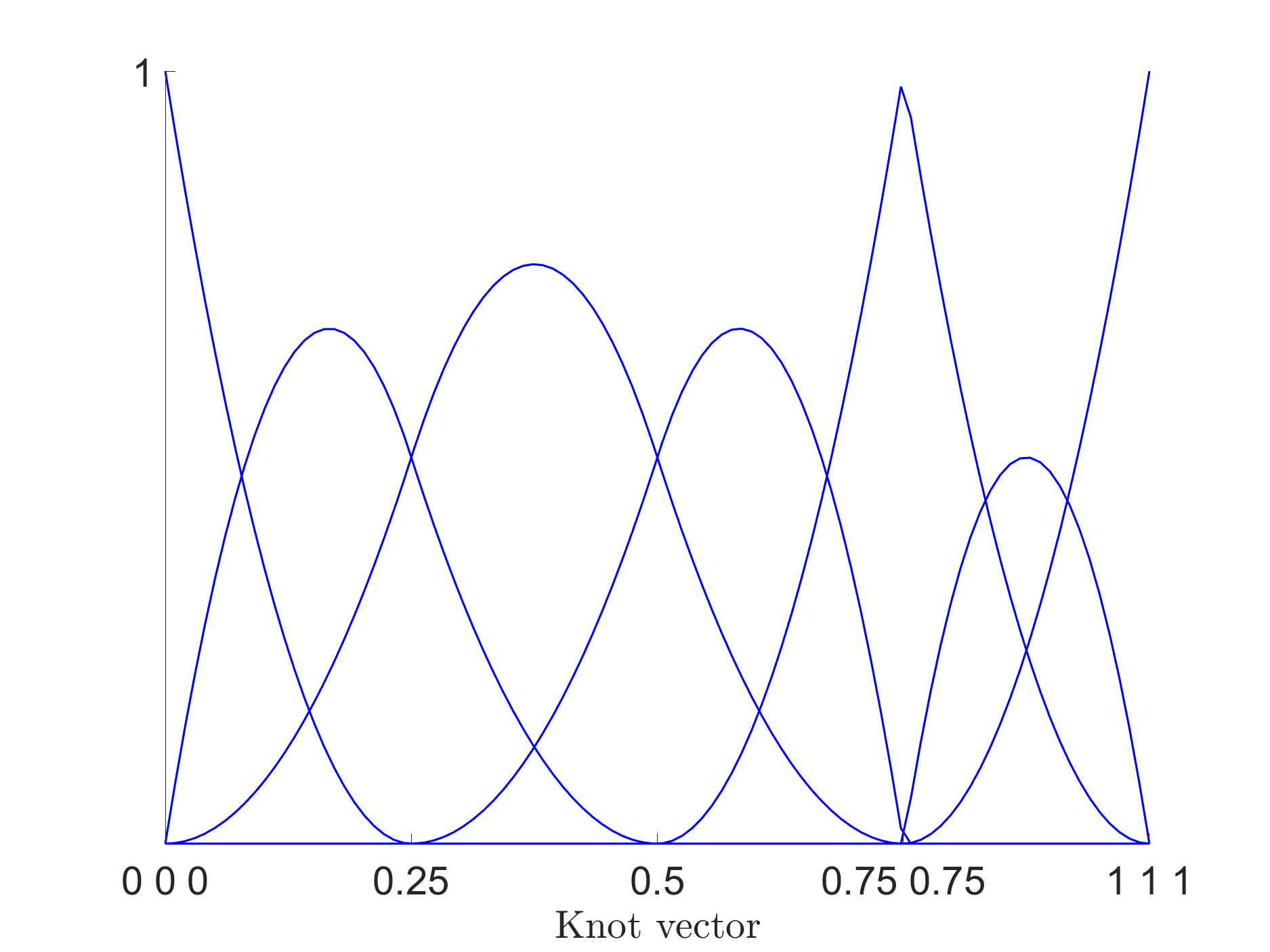}
    \end{minipage}
    \hfil
    \begin{minipage}{.49\textwidth}
    \includegraphics[scale=0.24]{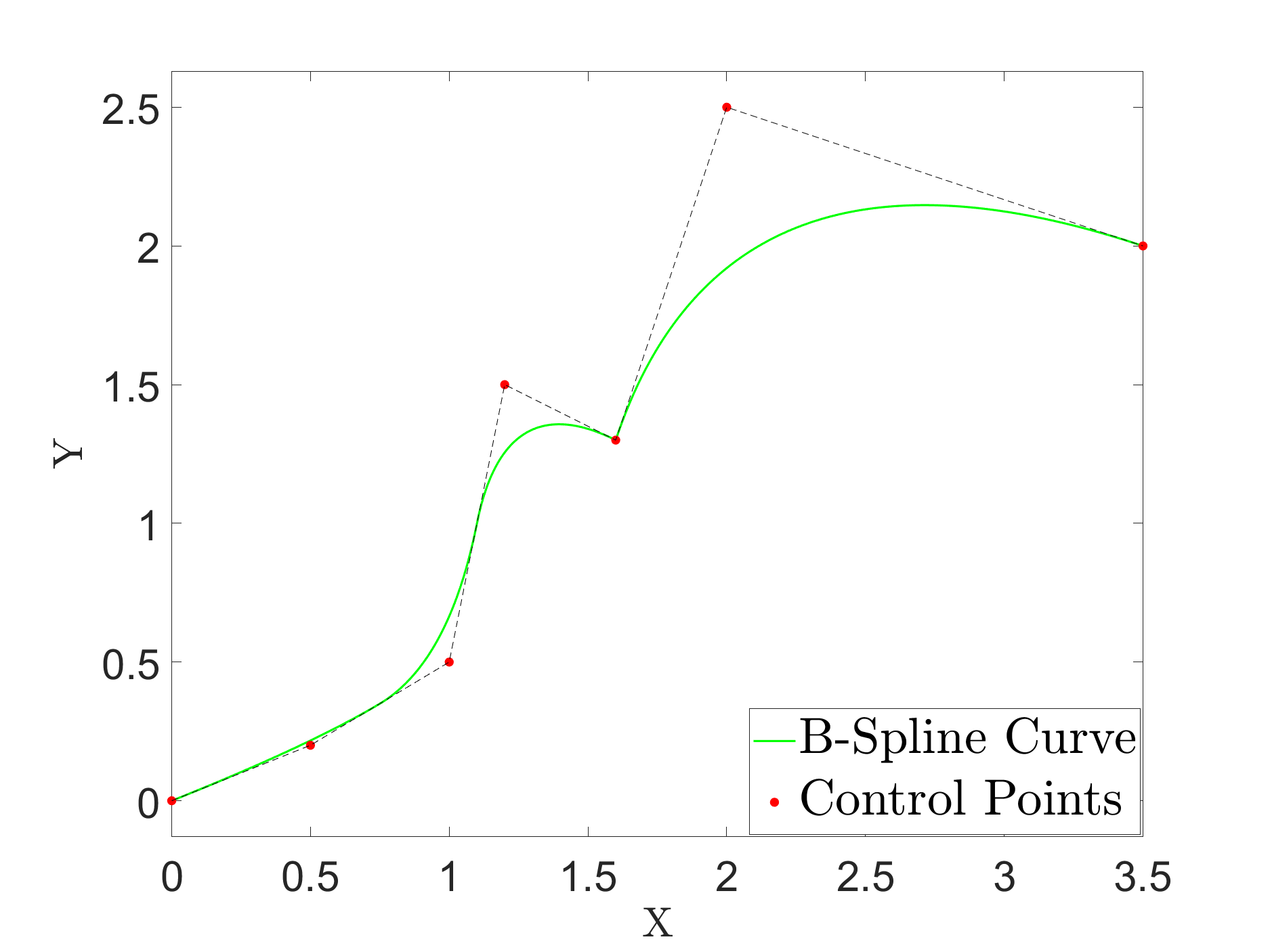}
    \end{minipage}
    \caption{Basis functions of a univariate quadratic B-spline and the corresponding curve. $C^0$ continuity at the $knot = 0.75$.}
    \label{fig:example_bspline}
\end{figure}

The multiplicity $k$ of a knot value can be found from the knot vector and defines the smoothness of the B-spline basis at this location, e.g., $C^{p-k}$. In the case of open knot vectors, the first and the last knot values have multiplicity $k = p + 1$, and the basis is interpolatory at the ends of its definition domain. A B-spline curve can be constructed as the linear combination of the basis functions.
\begin{equation}\label{three}
    C(\xi) = \displaystyle\sum_{i=1}^{n}N_{i,p}(\xi)\mathbf{P}_i,
\end{equation}
where $\mathbf{P}_{i}\in\mathbb{R}^{d}$ are the control points and equal the number of basis functions. The Fig.\ref{fig:example_bspline}, shows a representation of a B-spline with knot vector $\Xi = [0,0,0,0.25,0.5,0.75,0.75,1,1,1]$ and degree $p=2$.

Going into higher dimensions to construct multivariate B-splines, tensor product of univariate B-splines is formed.  Considering $\mathbb{R}^{2}$, the degrees $p$ and $q$ and knot vectors $\Xi = \{ \xi_1\leq...\leq \xi_{n+p+1}\}$ and $H = \{ \eta_1\leq...\leq \eta_{m+q+1}\}$, define each dimension of a surface. Thus, two sets of basis functions along with the control net ${\mathbf{P}_{i,j}}$ with $ i=1,2,...,n$, $j=1,2,...,m,$ describe the surface: 
\begin{equation}
    S(\xi,\eta) = \displaystyle\sum_{i=1}^{n}\displaystyle\sum_{j=1}^{m}N_{i,p}(\xi)M_{j,q}(\eta)\mathbf{P}_{i,j}
\end{equation}

\subsection{Standard B\'{e}zier extraction}
This section provides a short introduction to B\'{e}zier extraction \cite{borden2011isogeometric}. Unlike the standard finite element approaches, B-spline basis functions are defined globally (see Fig.\ref{fig:example_bspline}). B\'{e}zier extraction provides a mapping between the smooth spline space and a $C^0$ representation of B\'{e}zier elements. Those elements are defined by Bernstein polynomials, which are local, i.e., their support is non-zero only in a single element. Thus, conventional element-based FE routines can be enriched by splines in a straightforward manner.

B\'{e}zier extraction is generally based on knot insertion algorithms. The B\'{e}zier decomposition of the B-spline curve into piece-wise B\'{e}zier curves is obtained by increasing the multiplicity of each internal knot in $\Xi$, equals to the degree $p$. Starting by inserting a single knot $\mathbf{\hat{\xi}}$ into $\Xi$, so $\hat{\Xi} = \Xi \cup \hat{\xi}$, where $\hat{\xi} \in [\xi_s,\xi_{s+1}]$. This procedure continues until the desired multiplicity of the internal knots is achieved, resulting a new set of basis functions and a new set of control points that retain the geometry of the curve. Further details for knot insertion techniques can be found in \cite{hughes2005isogeometric}. The B-spline curve $C$ is described by a set of Bernstein polynomials $B_{i,p}$ and the Bernstein control points $\mathbf{\bar{P}}_i$
\begin{equation}
    C(\xi) = \displaystyle\sum_{i=1}^{\hat{n}} B_{i,p}(\xi) \mathbf{\bar{P}}_i 
\end{equation}

The control points $\mathbf{\bar{P_i}}$ can be computed as a linear combination of the original control points $\mathbf{P}_i$, which are represented by a matrix operator $\mathbf{E}$, the B\'{e}zier extraction operator. The same operator is used for the translation of the basis functions $N_{i,p}(\xi)$ to the Bernstein polynomials $B_{i,p}(\xi)$. The basis functions can be written in a vector format with $\mathbf{N}$ and $\mathbf{B}$, respectively. So, globally defined B-splines transformed into a basis of B\'{e}zier elements reads
\begin{equation}\label{eight}
    \mathbf{\bar{P} = {E}^T P} \\
    \mathbf{N = EB}.
\end{equation}

\subsection{Truncated hierarchical B-splines}
In this section, the hierarchical B-splines are presented, which are used to achieve local refinement and overcome the limitation of adaptivity due to the tensor product structure of splines. Then, the extension with the truncation mechanism is added to retain the partition of unity property \cite{giannelli2012thb,hennig2016bezier}.

\begin{figure}[H]
        \centering
        \includegraphics[scale=0.25]{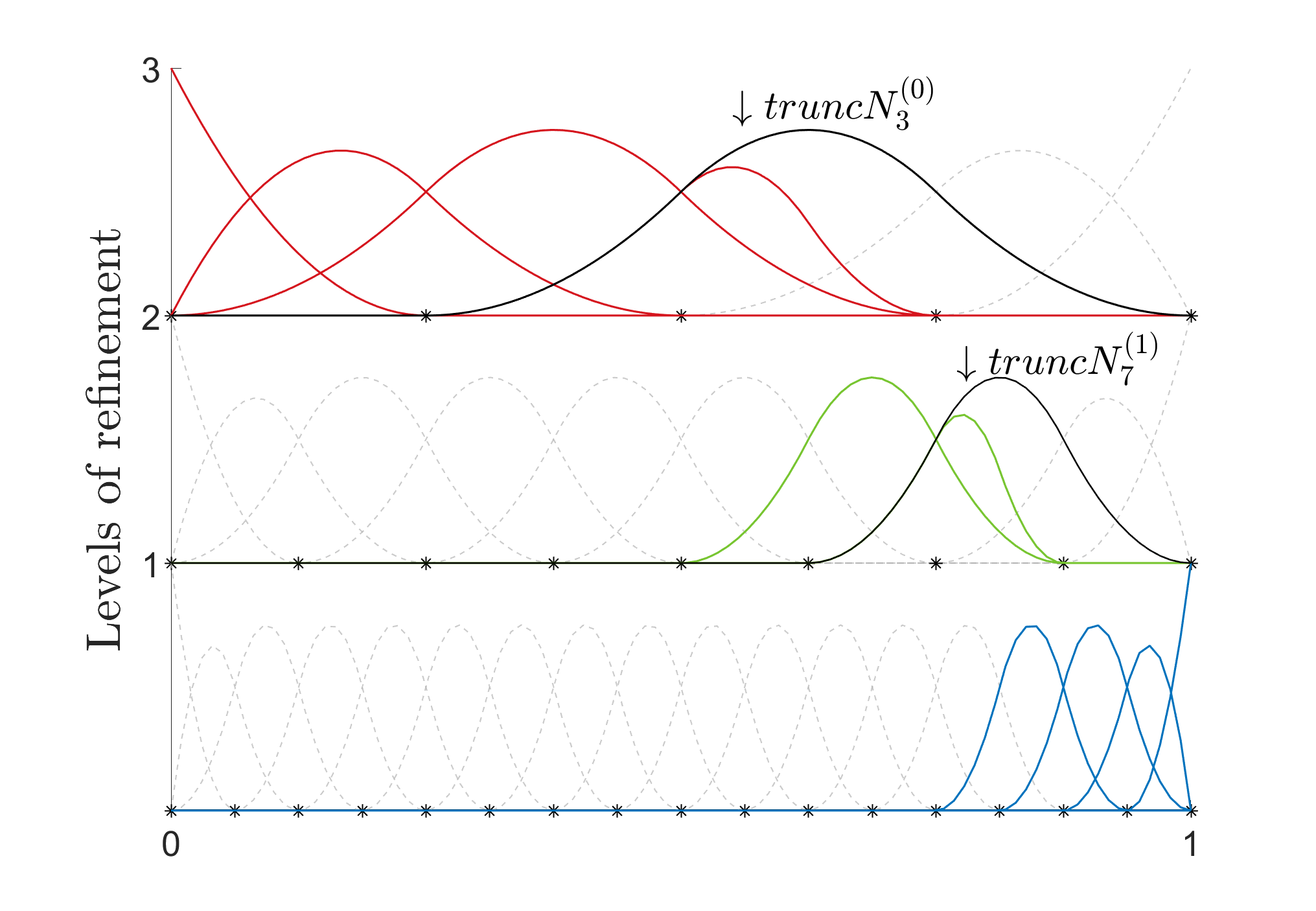}
        \caption{ Basis functions for the computational domain. The gray functions are the deactivated ones and the colored ones are the active ones. The $N_3^{(0)}$ and $N_7^{(1)}$ are truncated functions.}
        \label{fig:hb2spline}
    \end{figure}

Hierarchical B-splines \cite{kraft1997adaptive,vuong2011hierarchical} introduce a multi-level structure with different levels of refinement. Consider a sequence of nested spline spaces of univariate splines $\mathbb{S}^{(0)}(\Omega) \subset \mathbb{S}^{(1)}(\Omega) ... \subset  \mathbb{S}^{(\ell_{max})}(\Omega)$ with the same polynomial degree, where $\ell=0 \dots \ell_{max}$ are the levels of refinement. In this contribution, the different levels are constructed with the dyadic refinement technique, where each new level has $2d \cdot n_{el}$ elements ($d$ is the dimension and $n_{el}$ number of elements in $\ell-1$). In Fig. \ref{fig:hb2spline}, a univariate quadratic B-spline with two additional levels of refinement is depicted. The geometry and the parametric representation do not change under the refinement process, but the relation between the levels is happening by an operator called, subdivision matrix $\mathbf{S}$. The parent functions of the level $\ell$ are a linear combination of fine-scale functions (children functions) of level $\ell + 1$ 
\begin{equation}\label{Submat}
        N(\xi)_{i,p} = \displaystyle\sum_{j=1}^{\hat{n}} \mathbf{S_{j,i}} \hat{N}_{j,p}(\xi). 
\end{equation}

Each nested knot vector $\Xi^{\ell}$ defines a new set of B-spline basis functions $N_{j,p}^{\ell}(\xi)$. So, the multi-level spline space is defined as in \cite{marussig2018improved}
\begin{equation}\label{ten}
\begin{split}
    MS(\Omega) := \text{span}\{ {N^{(\ell)}_i \text{for i} \in \eta^{(\ell)}_A, \ell = 0,1,\hdots,\ell_{max}}\} \\
    \eta^{(\ell)}_A\ \text{: active},\eta^{(\ell)}_D \text{: deactivate},  \eta^{(\ell)}_A \cup \eta^{(\ell)}_D = \eta^{(\ell)}. 
\end{split}
\end{equation}

 Therefore, the refinement process starts by deactivating parent functions of the corresponding element from the coarse level $\ell$ and activating their children on level $\ell + 1$, until the desired adaptivity is achieved, see Fig. \ref{fig:hb2spline}. The selection of the coarse and fine elements results from error estimator algorithms or manual selection of the preferred elements. This information is stored in the subdivision matrices \cite{marussig2018improved,kiss2015theory}.
 
 The truncated basis was first time introduced in \cite{giannelli2012thb}. The truncation mechanism preserves all the nice properties of hierarchical B-splines, such as linear independence, local support, and non-negativity. In addition, truncated hierarchical B-splines (THB-splines) have smaller support and retain the partition of unity property of traditional B-splines. The truncated basis functions are defined as
\begin{equation}\label{trunc}
    trunc N^{\ell}_i(\xi) = \displaystyle\sum_{j \in \eta^{\ell+1}_D} S^{(\ell)}_{ji} N^{(\ell+1)}_j(\xi).
\end{equation}

In general, it is straightforward to identify the truncated B-spline functions. Let's assume an evaluation of an element on the finest level $\ell + 1$. The basis functions that describe the element are $N^{\ell+1}_i(\xi)$, with $i=s^{\ell+1}-p \hdots s^{\ell+1}$, where the $s^{\ell+1}$ is the knot span of the element in that level. This element holds truncated B-splines on the previous level, $\ell$, only if at least one of its $N^{\ell+1}_i(\xi)$ is inactive (see Fig. \ref{fig:hb2spline}).
\section{Multi-level B\'{e}zier extraction}   

\begin{figure}[H]
    \begin{minipage}{.32\textwidth}
    \includegraphics[scale=0.25]{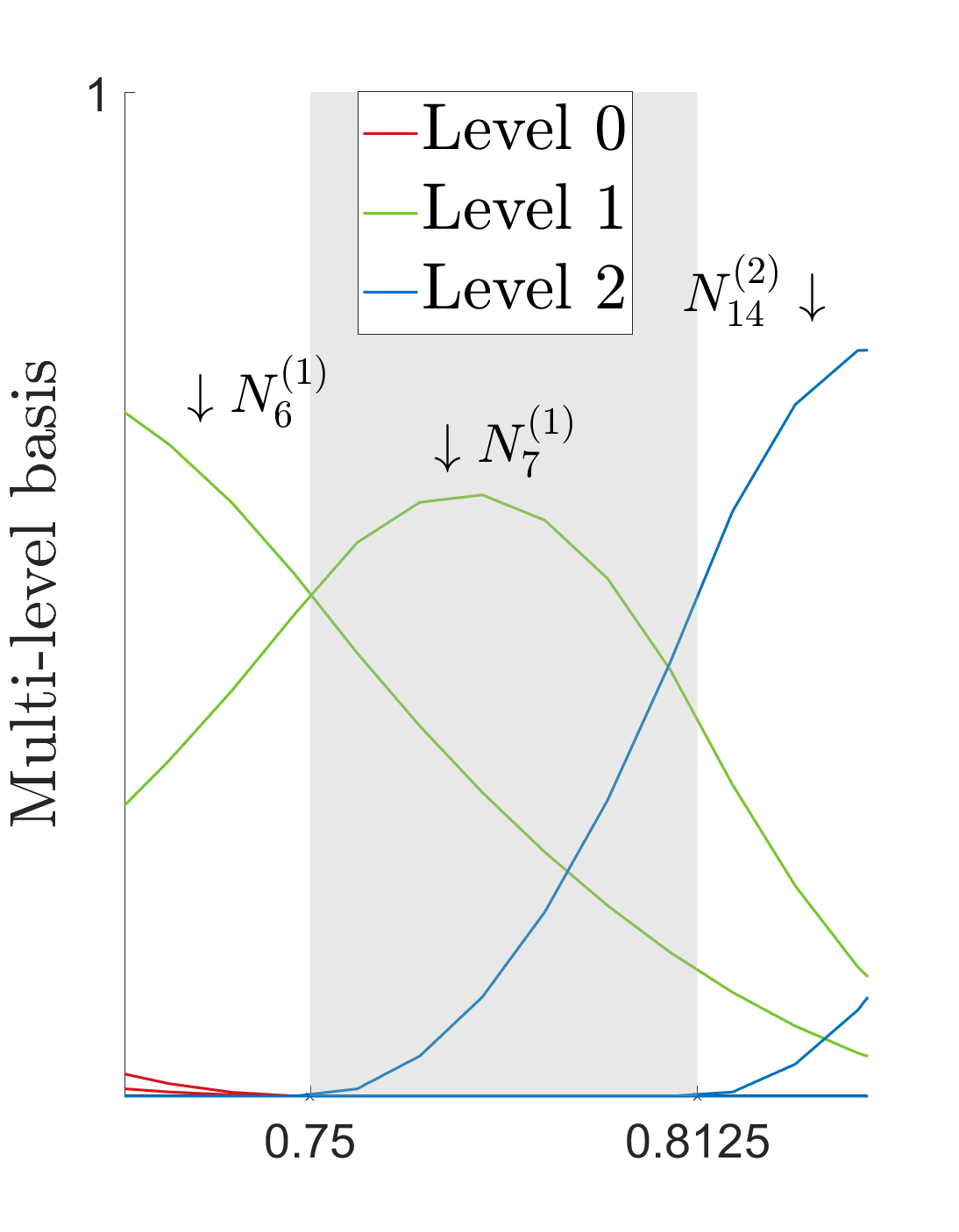}
    \end{minipage}
    \hfil
    \begin{minipage}{.32\textwidth}
    $\begin{bmatrix}
                N_{6}^{(1)} \\
                N_{7}^{(1)} \\
                N_{14}^{(2)}
            \end{bmatrix} = \mathbf{M}^{e_3} \mathbf{E}^{e_3}  \begin{bmatrix}
            B_{0}\\
            B_{1}\\
            B_{2} 
        \end{bmatrix}$
    \end{minipage}
    \hfil
    \begin{minipage}{.32\textwidth}
    \includegraphics[scale=0.25]{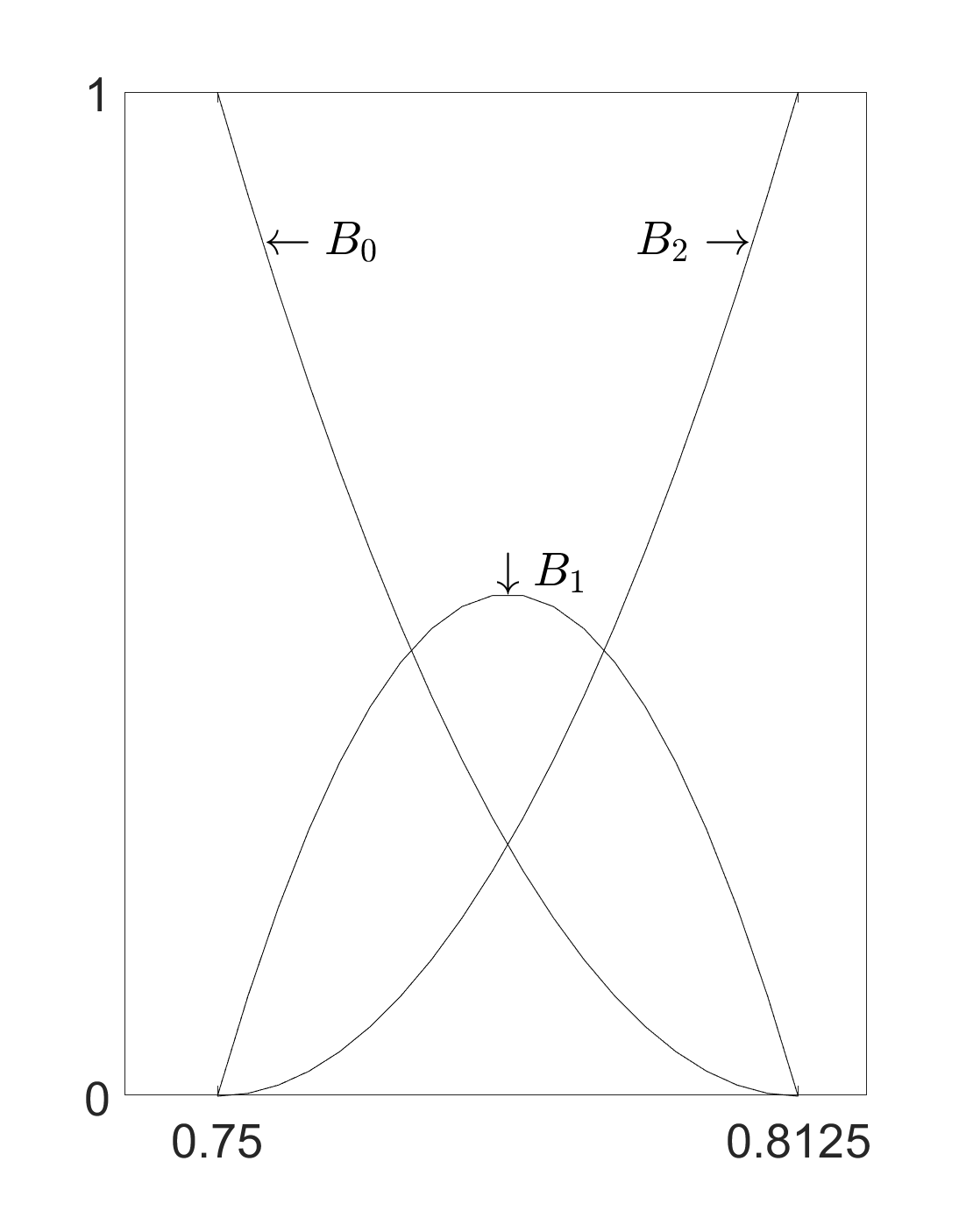}
    \end{minipage}
    \caption{Translate from hierarchical spline space to local Bernstein polynomials.}
    \label{fig:translation}
\end{figure}

        Having a sequence of nested spline spaces into a hierarchical structure, every basis function of level $\ell$ can be expressed as a linear combination of basis functions of level $\ell+1$, with the subdivision matrix \eqref{Submat}. The global multi-level extraction operator, $\mathbf{M}^{glob}_{\ell}$, can be defined by applying a sequence of nested spline spaces, where the goal is to extract the hierarchical functions supporting an element at a specific level. For levels $\ell=0,..,\ell_{max}$ the $\mathbf{M}^{glob}_{\ell_{max}}$ can be obtained by joining the rows, corresponding to the active basis, of the refinement operators $\mathbf{S}^{(\ell,\ell_{max})}$. More details for constructing the global multi-level extraction operator can be found in \cite{d2018multi}. Next, localization of the global multi-level extraction to each element of the domain leads to a local element-by-element approach. By selecting the appropriate columns and rows associated with the basis functions of the element, the extraction operator effectively captures the hierarchical functions. The multi-level extraction operator is always of the same level as the element. All the different matrices, $\mathbf{M}^{loc}_{e}$, correspond to each element of the hierarchical spline space, $e=0,...,n_{el}$. They can be combined directly with the B\'{e}zier extraction operator, resulting the multi-level B\'{e}zier extraction technique (see Fig. \ref{fig:translation}). This creates a direct map $\mathbf{C^e}$ from a standard set of reference basis functions equal for each element (Bernstein basis, $\mathbf{B}$) to the multi-level local basis (hierarchical basis). 

    \begin{equation}\label{MB}
        \begin{split}
            H^e &= \mathbf{M}^{loc}_{e} \mathbf{E}^{loc}_{e} \mathbf{B}\\
            &= \mathbf{C}^{e} \mathbf{B}
        \end{split},\quad \textnormal{with} \quad
        \mathbf{B} = \begin{bmatrix}
                B_{0}\\
                B_{1}\\
                \vdots\\
                B_{p} 
        \end{bmatrix}
    \end{equation}

As the hierarchical basis functions are mapped into a local approach with elements with a standard set of basis functions, the calculations into the B\'{e}zier elements are performed. This can happen using the existing FE algorithms, and there is no need for separate algorithms to handle the globally defined spline basis functions.

\section{Numerical example}
\begin{figure}[H]
    \begin{minipage}{.49\textwidth}
    \includegraphics[scale=0.33]{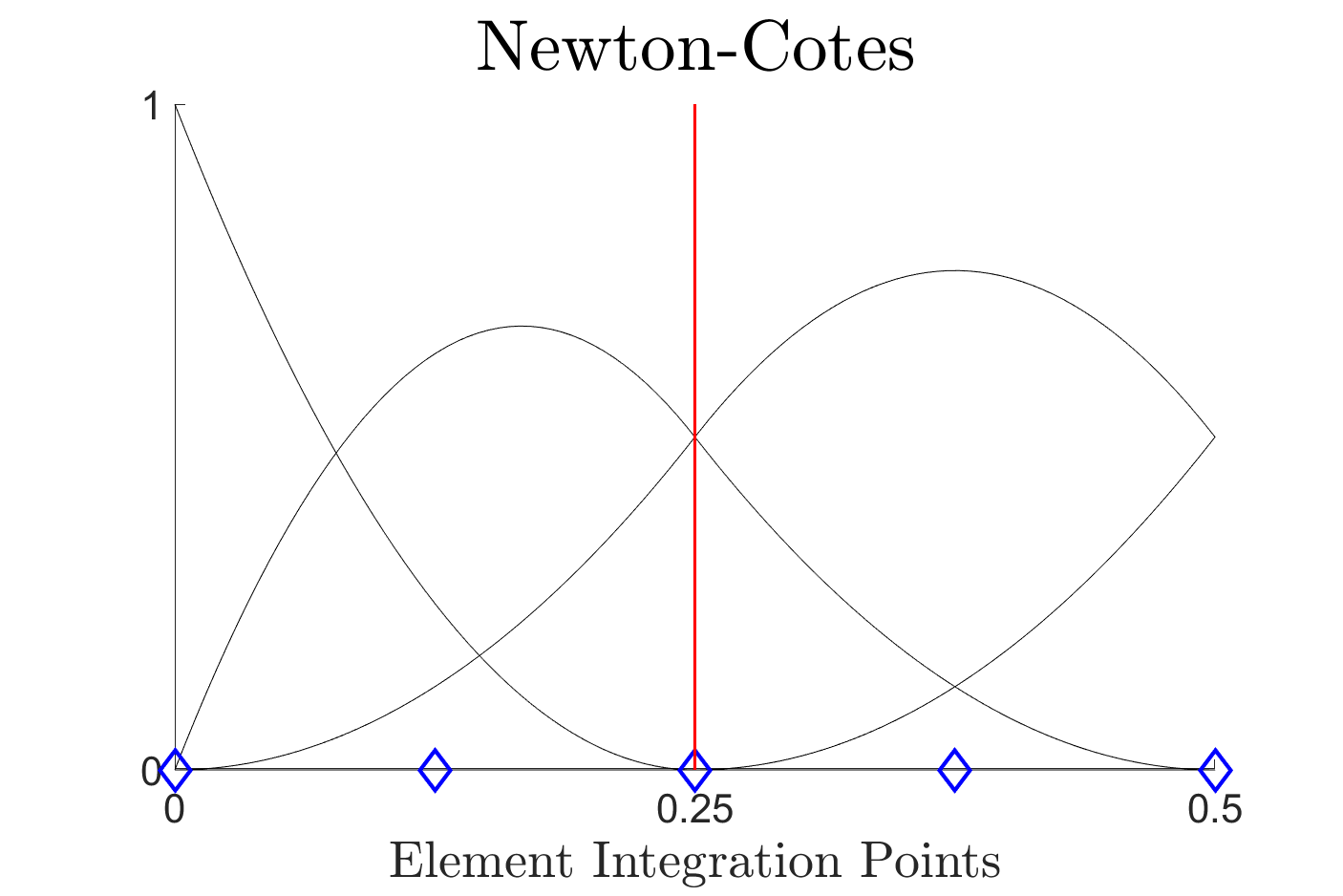}
    \end{minipage}
    \hfil
    \begin{minipage}{.49\textwidth}
    \includegraphics[scale=0.33]{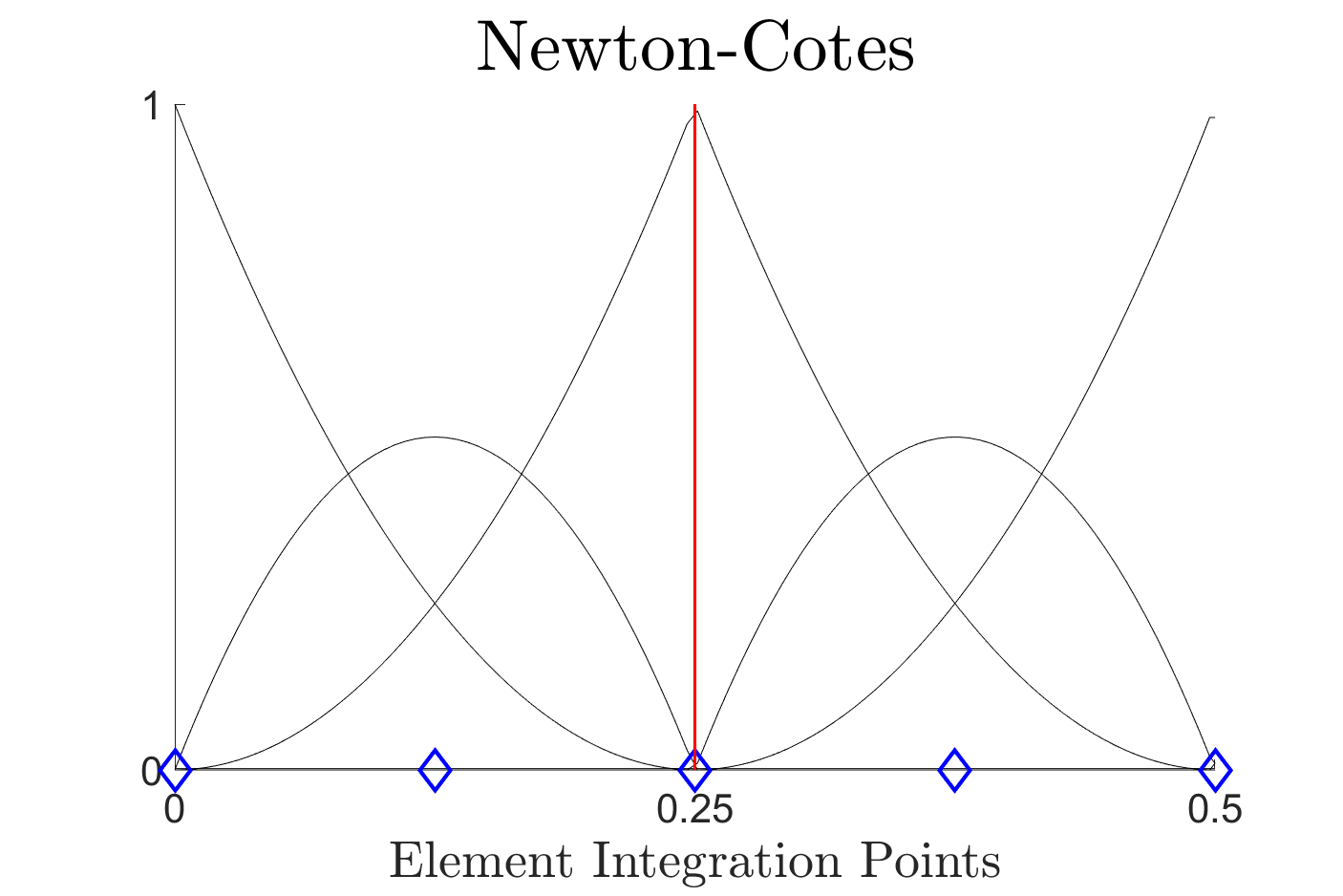}
    \end{minipage}
    \caption{Newton-Cotes integration points appear in the boundary of the elements and can not be evaluated correctly due to the Cox-de Boor formula. B\'{e}zier extraction, in the right figure, overcomes this limitation with a standard set of Bernstein polynomials.}
    \label{fig:example_NC}
\end{figure}

This contribution uses GeoPDEs \cite{garau2018algorithms,vazquez2016new}, an open-source FE solver for IGA, based on Matlab/Octave. Let us first highlight an advantage of B\'{e}zier extraction over the direct use of B-splines basis functions. The problem refers to the evaluation of points at the boundaries of the elements. Due to the Cox-de Boor formula \eqref{Cox}, boundary points are generally associated with the right element of the domain, which can lead to misinterpretations. Consider the numerical integration of an element using a Newton-Cotes quadrature rule. These rules contain the boundary points, and based on \eqref{Cox}, the contribution of the last point would be associated with the next element instead of the current one. This can be easily overcome with the use of B\'{e}zier extraction. An example to illustrate the problem is in Fig. \ref{fig:example_NC}, where the Newton-Cotes quadrature rule is applied for the integration of the elements in a univariate problem, instead of the Gaussian quadrature method. B\'{e}zier extraction should be applied to evaluate accurately. In this case, improving the results is not the goal of Newton-Cotes, but to verify the problem of spline-based algorithms using Cox-de Boor formula.

\subsection{Unit square boundary value problem}
\begin{figure}[H]
    \centering
    \includegraphics[scale=0.14]{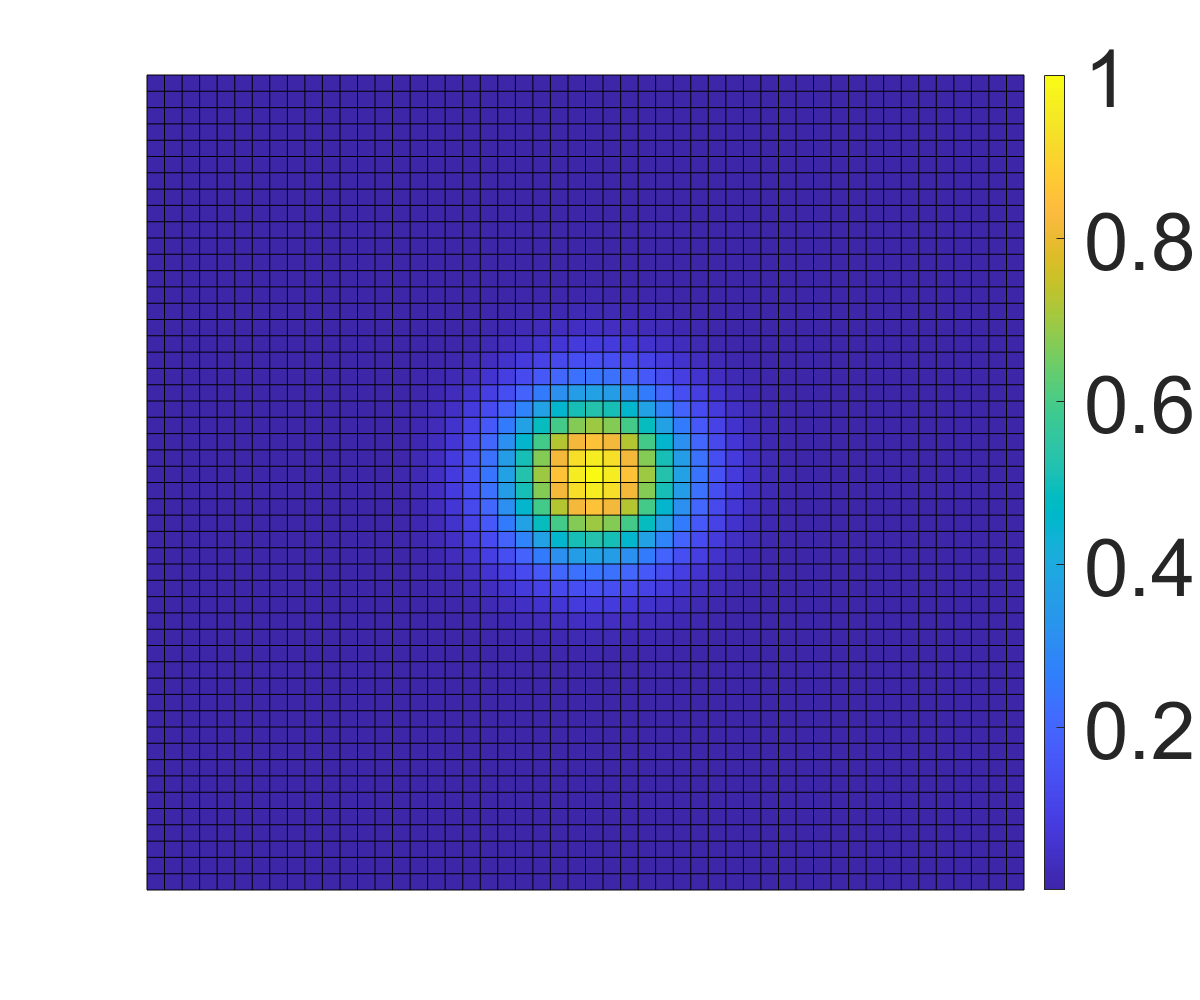}
    \caption{Exact solution of the unit square boundary value problem.}
    \label{fig:example_exactsol}
    \end{figure}

\begin{figure}[H]
    \begin{minipage}{.24\textwidth}
        \includegraphics[scale=0.26]{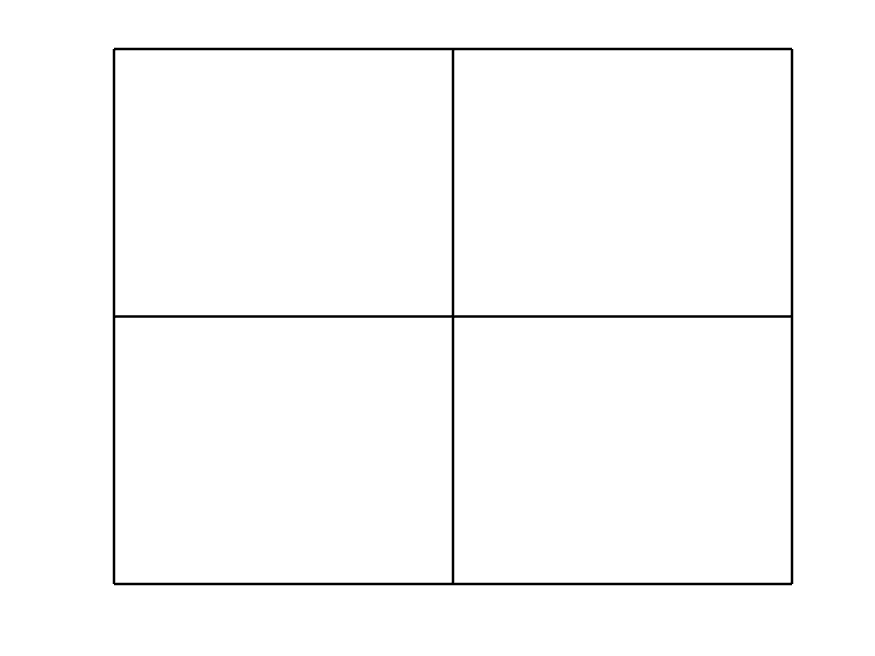}
    \end{minipage}
    \hfill
    \begin{minipage}{.24\textwidth}
        \includegraphics[scale=0.26]{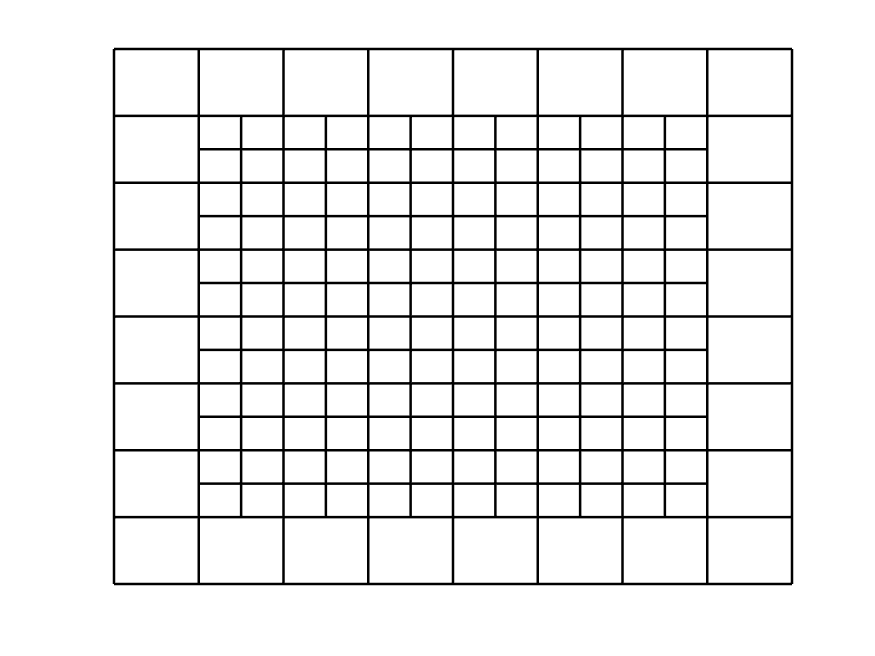}
    \end{minipage}
    \hfill
    \begin{minipage}{.24\textwidth}
        \includegraphics[scale=0.26]{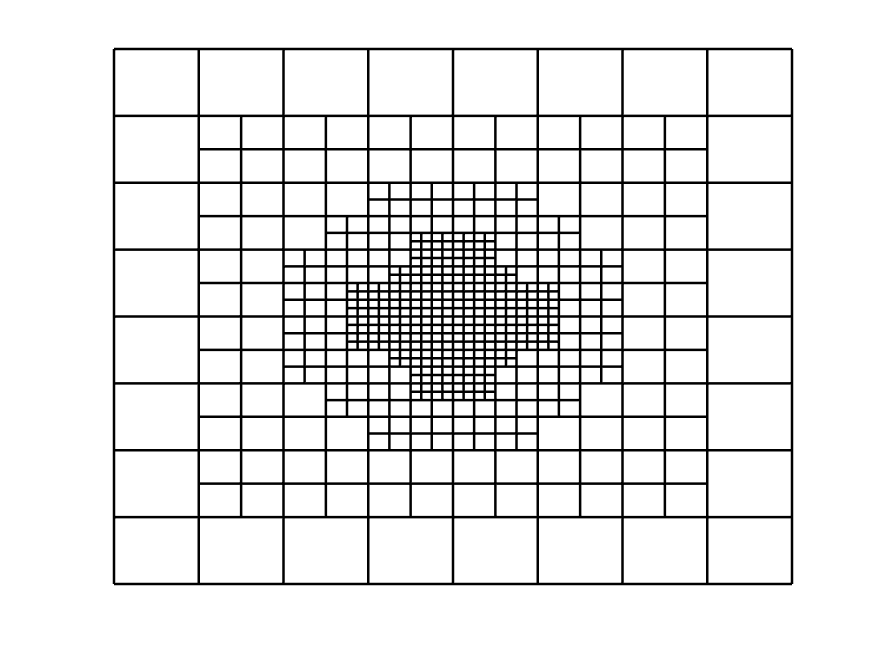}
    \end{minipage}
    \hfill
    \begin{minipage}{.24\textwidth}
        \includegraphics[scale=0.26]{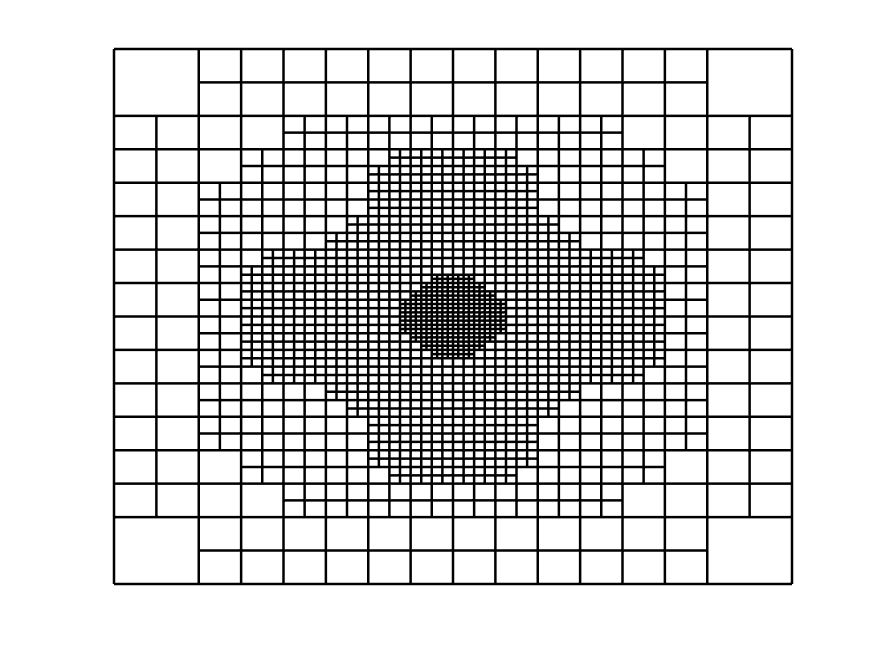}
    \end{minipage}
    \caption{Mesh through iterations of local refinement with THB-splines.}
    \label{fig:ref}
\end{figure}

The numerical example for this work is a second-order elliptic boundary value problem in a unit square domain. The functions $f$, $g$ and $h$ are chosen from the exact solution (see Fig. \ref{fig:example_exactsol}) of the problem $u_{ex} = \exp{(-C \cdot ((x-0.5)^2 + (y-0.5)^2))}$, where $C$ is constant.
    \begin{equation}
        \begin{cases} \Delta u= f & \textnormal{in}\:\: \Omega\\ u = g & \textnormal{on}\:\: \Gamma_D \\ \triangledown u\cdot \mathbf{n} = h & \textnormal{on}\:\: \Gamma_N \end{cases}, \\
        \Omega=\left[0, 1\right]^{2} \: \& \: {\Gamma_N \cup \Gamma_D} = \partial{\Omega}
    \end{equation} 
    For the calculation, two types of THB-splines are assumed, one quadratic and one cubic. In both cases, global and local refinement techniques are compared. For the local refinement, the refined mesh is shown in Fig. \ref{fig:ref}. Now, as for the multi-level B\'{e}zier extraction, the goal is to approach the error of the local refinement that GeoPDEs is resulting in. Calculating the multi-level B\'{e}zier extraction operator $\mathbf{C}^{e}$, as described in Section 3, and then calculating the stiffness matrix of a single B\'{e}zier element, the stiffness matrix of an element in the hierarchical structure can be derived.

    \begin{equation}\label{fourteen}
        \begin{split}
            \mathbf{K}_{B-spline,e} = \mathbf{C}^{e}  \mathbf{K}_{Bezier}  (\mathbf{C}^{e})^T \\
            \mathbf{K}_{global} = \displaystyle\sum_{i=1}^{n_{el}} \mathcal{T}(\mathbf{K}_{B-spline,i}^e)
        \end{split}
    \end{equation}

\begin{figure}[H]
\begin{minipage}{.48\textwidth}
    \includegraphics[scale=0.24]{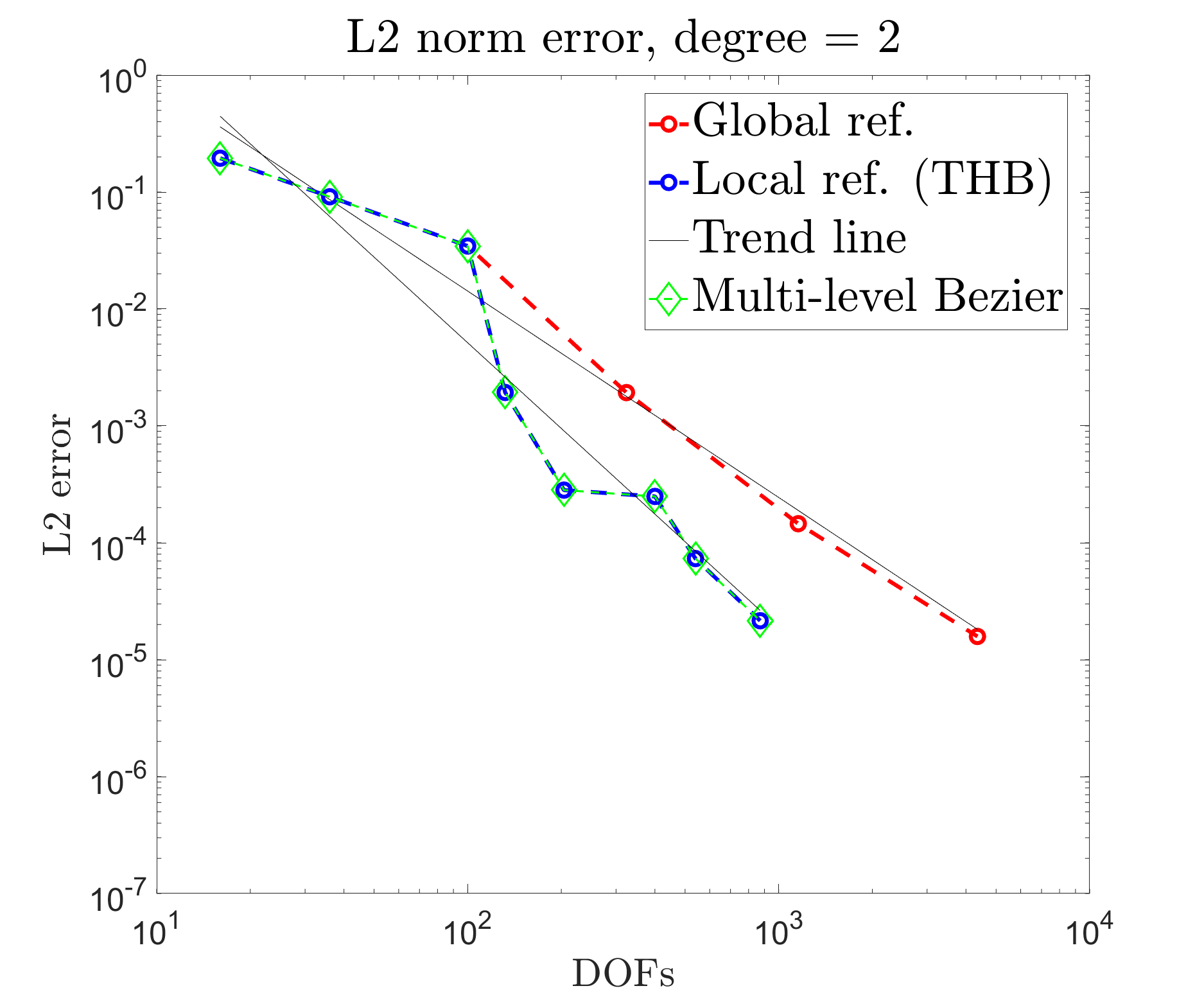}
\end{minipage}
\hfil
\begin{minipage}{.48\textwidth}
    \includegraphics[scale=0.24]{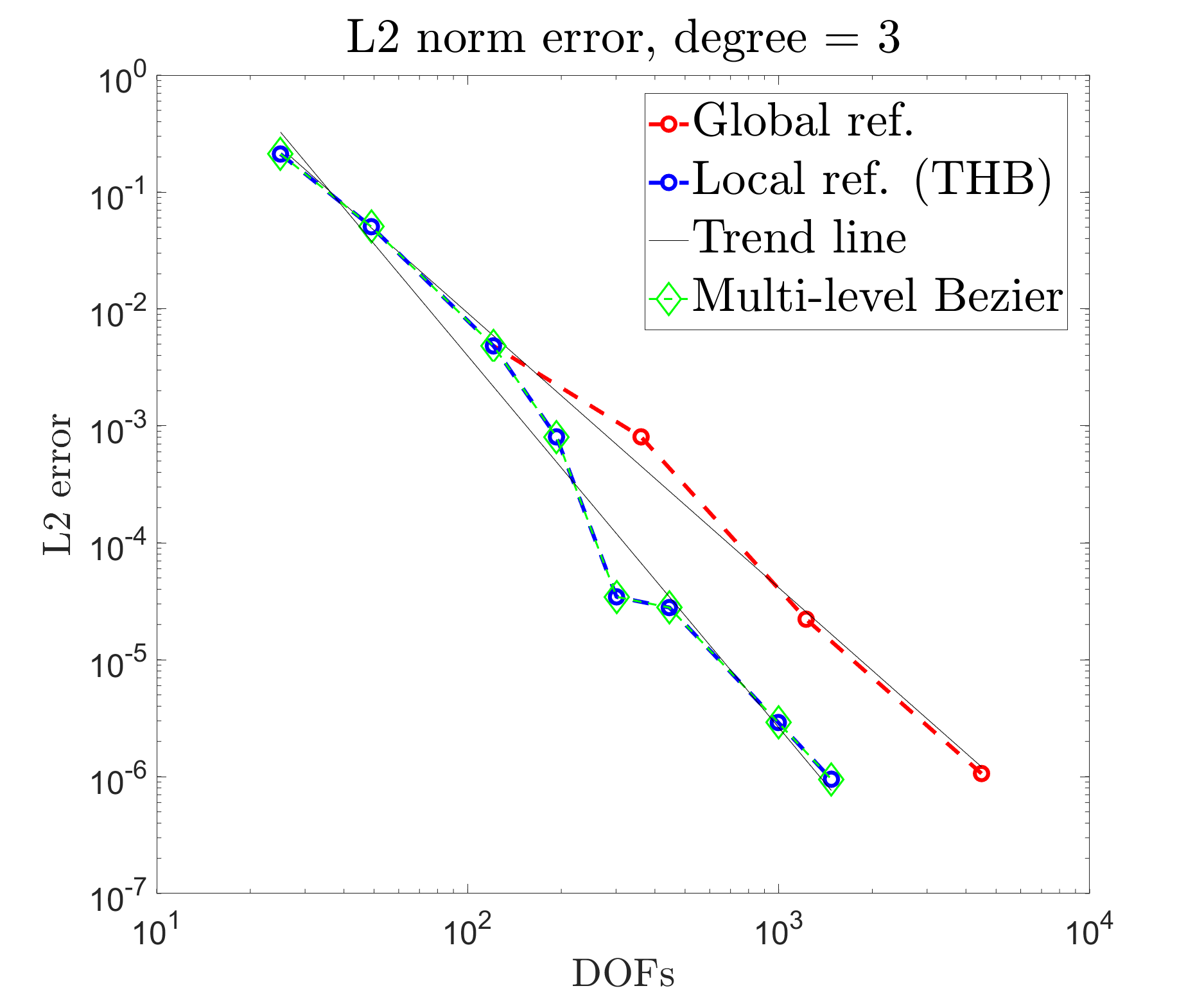}
\end{minipage}
\caption{L2 norm error. Global refinement, local refinement, and multi-level B\'{e}zier extraction on the local refinement. On the left are quadratic splines, and on the right are cubic splines. The green line of multi-level B\'{e}zier extraction matches exactly the blue line of the local refinement from the patch-wise approach.}
\label{fig:results}
\end{figure}

Gathering all the element-wise stiffness matrices, the stiffness matrix for the whole multi-level spline space is obtained. Fig. \ref{fig:results} shows the convergence of the multi-level B\'{e}zier extraction and the local refinement of GeoPDEs. 

When comparing the values of the graphs in both cases, the results of multi-level B\'{e}zier extraction fit the IGA-based algorithm. This verifies that by just evaluating a single B\'{e}zier element, the stiffness matrix can be directly mapped into a hierarchical structure and evaluate multi-level spline spaces.
\section{Conclusion}
This study demonstrates that B\'{e}zier extraction can be extended into a multi-level structure of truncated hierarchical B-splines (THB-splines), resulting the multi-level B\'{e}zier extraction. The calculation over B\'{e}zier elements is sufficient to extract results for hierarchical spline spaces which perform local refinement. The only calculation that needs to be derived is the multi-level B\'{e}zier extraction operators for each element of the domain and the standard B\'{e}zier extraction operator. Then those operators are applied to B\'{e}zier elements. The implementation of the isogeometric analysis concept into existing finite element solvers can be employed. Additionally, this general approach is not restricted to THB-splines but can be compatible with any nested space.
\section{Ackowledgments}
The work is supported by the joint DFG/FWF Collaborative Research Centre CREATOR (DFG: Project-ID 492661287/TRR 361; FWF: 10.55776/F90) at TU Darmstadt, TU Graz and JKU Linz.
\vspace{\baselineskip}

\end{document}